\documentclass[12pt]{article}

\usepackage{tikz,pgfplots}
\usetikzlibrary{trees,arrows}
 
\usepackage{amsmath}
\usepackage{graphics}
\usepackage{epsfig}

\usepackage{amssymb}
\usepackage{amscd} 
\usepackage[all]{xy}
\usepackage{graphicx}
\usepackage{amsfonts, amsmath, amsthm, amssymb} 

\title{Quantum Coin Flipping, Qubit Measurement  and Generalized Fibonacci Numbers}

\author{Oktay K. Pashaev\\Department of Mathematics\\ Izmir Institute of Technology\\Urla - Izmir, 35430, Turkey}

\begin{document}

\maketitle

\begin{abstract}
The problem of Hadamard quantum coin measurement  in  $n$ trials, with arbitrary number of repeated consecutive  last states is formulated in terms of Fibonacci sequences for duplicated states, Tribonacci numbers for 
triplicated states and $N$-Bonacci numbers for arbitrary $N$-plicated states. The probability formulas  for arbitrary position of  repeated states are derived in terms of Lucas and Fibonacci numbers.
For generic qubit coin, the formulas  are expressed by Fibonacci and more general, $N$-Bonacci polynomials in qubit probabilities. The generating function for probabilities, the Golden Ratio limit of these probabilities
and Shannon entropy for corresponding states are determined. By generalized Born rule and universality of $n$-qubit measurement gate, we formulate problem in terms of generic $n$-qubit states and construct 
projection operators in Hilbert space, constrained on the Fibonacci tree of the states. The results are 
generalized to qutrit and qudit coins, described by generalized Fibonacci-$N$-Bonacci sequences. 
\end{abstract}


\section{Introduction}

The Fibonacci numbers have been known from ancient times as
"nature’s numbering system", and have applications to the grows
of every living thing, from natural plants (branches of trees,
arrangement of leaves) to human proportions and architecture
(the Golden Section $\varphi = \frac{1 + \sqrt{5}}{2} \approx 1.6$).
Quantum calculus of Fibonacci numbers and Fibonacci divisors, as $q$-numbers with Golden Ratio bases  was developed in our papers
\cite{p1}, \cite{p2}. Some applications of this calculus, as descriptive of $n$-qubit states, based on Binet formula for coherent qubit  states were derived. Normalization constants and probability of measurement in these states are determind by Fibonacci numbers and Fibonacci divisors. In particular, it was shown that for two qubit state
with $n=2$, and arbitrary odd Fibonacci divisor $F_n^{(k)}$, the level of entanglement in terms of concurrence is expressible in terms of Lucas numbers $L_k$ only, as decreasing function of $k$ \cite{p2}.

In the present paper we describe the measurement problem for qubit states with repeated identical states, which is naturally related with Fibonacci numbers and their generalizations to Tribonacci and N-Bonacci cases.
For maximally random qubit state, as the Hadamard quantum coin state, this problem is the quantum analog of the classical coin flipping. The last one is the known from XVIII century problem, formulated by A. De Moivre in his book on probability theory "The doctrine of chanses" \cite{demoivre}. Some useful references on coin flipping problem and corresponding classical dynamical system can be found in \cite{kennedi}.
Here we are going to work with quantum coin as two-level quantum system, representing the qubit unit of quantum information. This quantum coin can be repreneted by spin one-half, so that the measurement of this spin in $z$-component, plays the role of quantum coin flipping \cite{aharonov} . 
The quantum coin flipping has application as a protocol to encrypt messages for secure quantum communication.
A single-qubit coin determines one-dimensional \cite{aharonov} and alternate two-dimensional \cite{franco} quantum walks. The Hadamard gate represents simplest fair quantum coin and corresponding Hadamard quantum walk \cite{quantumwalk}.  The interesting question is a relation between classical and quantum coins. From one side in \cite{manko}, the quantum states are mapped to the classical coin probabilities by replacing A. Einstein's sentence "God does not play dice" by statement "God plays coins". From another side, according to observation by A. Albrecht and D. Phillips,
the outcome of coin flip is truly quantum measurement, and it is really a Schr\"odinger cat, so that 50-50 outcome of a coin toss may be derived from quantum physics      \cite{albrecht}. 

In the present work we formulate quantum version of classical coin flipping problem, as quantum measurement problem for quantum coin. In Section 2 we introduce maximally random qubit state as a quantum coin. In Section 3 we describe measurement of this qubit state as Hadamard coin. By generalized Born rule and universality of $n$-qubit measurement gate, we formulate problem in terms of generic $n$-qubit states.
We show that for duplicated states,  the Fibonacci sequence of numbers determines allowed configurations of states and corresponding probabilities. It turns out that the Fibonacci tree for computational quantum states becomes identical to classical Fibonacci rabbit problem and corresponding tree.
For triplicated states and generic N-plicated states, these configurations are determined by sequences of Tribonacci and $N$-Bonacci numbers. Generic qubit coin measurement is derscribed in Section 4. In Section 5 we consider arbitrary position 
of duplicated, triplicated and N-plicated states and corresponding probabilities. Section 6 is devoted to generating functions for probabilities, the Golden Ratio in measurements and Shannon entropy for the states.
Projection operators on quantum tree states and corresponding probabilities are subject of Section 7. In Section 8 we treat quantum qutrit coin with three output states of measurement and generalized Fibonacci and Tribonacci numbers. In Section 9 we consider most general case of qudit coin and coresponding generalized Fibonacci type sequences.

 \section{Maximally random qubit state as quantum coin}

For one qubit state in computational basis
\begin{equation}
| \psi\rangle = c_0 |0\rangle + c_1 |1\rangle,\label{qubit}
\end{equation}
where $|c_0|^2 + |c_1|^2 =1$, probabilities of measurement the states $|0 \rangle$
and $| 1 \rangle$ are $p_0 =|\langle 0| \psi \rangle|^2 $ and  $|\langle 1| \psi \rangle|^2$, correspondingly. From this point of view, the qubit can be considered as a random variable state with two outcome states and corresponding probabilities, $p_0 + p_1 =1$. 
The Shannon entropy for these probabilities
\begin{equation}
S = -p_0 \log_2 p_0 - -p_1 \log_2 p_1 =  -|c_0|^2 \log_2 |c_0|^2 - |c_1|^2 \log_2 |c_1|^2,
\end{equation}
rewritten in the form
\begin{equation}
S_{}  =  -|\langle 0| \psi \rangle|^2 \log_2 |\langle 0| \psi \rangle|^2 - |\langle 1| \psi \rangle|^2 \log_2 |\langle 1| \psi \rangle|^2
\end{equation}
is the natural measure of the uncertainty in the result of a measurement  \cite{deutsch}, quantifying the deficiency in the information, which 
the outcome state $| \psi\rangle$ gives about further measurements. The entropy depends on the basis states and is changing with transformation of the basis.
It takes maximal value $S = 1$ for $p_0 = p_1 = \frac{1}{2}$ or $|c_0| = |c_1| = \frac{1}{\sqrt{2}}$,  giving    the Hadamard type qubit states
\begin{equation}
| \varphi\rangle = \frac{1}{\sqrt{2}} (|0\rangle + e^{i\varphi} |1\rangle).\label{quantumcoin}
\end{equation}
The Hadamard states
\begin{equation}
| \pm\rangle = \frac{1}{\sqrt{2}} (|0\rangle \pm |1\rangle),
\end{equation}
are generated by the Hadamard gate, $| +\rangle = H | 0\rangle, | -\rangle = H | 1\rangle$
and
correspond to $\varphi=0$ and $\varphi = \pi$ respectively. Since probabilities to measure states $|0\rangle$ and $|1\rangle$ are equal $p_0 = p_1 = \frac{1}{2}$, the qubit state (\ref{quantumcoin})
is naturally to call as the quantum coin. The states of this quantum coin, as maximally random states,  belong to the unit circle on equator of the Bloch sphere, the midway between $|0 \rangle$ and $|1\rangle$ states,
\begin{equation}
| \theta = \frac{\pi}{2}, \varphi\rangle = \frac{1}{\sqrt{2}} (|0\rangle + e^{i\varphi} |1\rangle).\label{qcoin}
\end{equation}
The whole set of states $|\varphi \rangle$ in (\ref{quantumcoin}) can be generated by the Phase gate  $R(\varphi)$,
 applied to the Hadamard state $|+\rangle$, which gives rotation of the unit circle

\begin{equation}   |+\rangle  \hskip 0.5cm \line(1,0){50}\fbox{\rule[-.3cm]{0cm}{1cm} R($\varphi$)}  \line(1,0){50}  \hskip 0.5cm  |\varphi\rangle
\label{rotation}\end{equation}

The classical coin could exist in two sates, but not in superposition of these states. The generic superposition of two quantum coins (maximally random qubits), is not maximally random state and is not fair quantum coin anymore. Indeed,
for the state
\begin{equation}
\alpha |+\rangle + \beta |-\rangle = \frac{\alpha + \beta}{\sqrt{2}}|0\rangle + \frac{\alpha - \beta}{\sqrt{2}}|1\rangle,\label{superposition}
\end{equation}
where, $|\alpha|^2 + |\beta|^2 =1$, probabilities are not equal
\begin{equation}
p_0 = \frac{1}{2} + \Re (\alpha \bar\beta),\,\,\,\,p_1 = \frac{1}{2} - \Re (\alpha \bar\beta) \label{ab}
\end{equation}
and the state is not maximally random. These type of states we call as the generic qubit coin states.
To make the superposition maximally random, in (\ref{ab}) we have to choose $p_0 = p_1 = \frac{1}{2}$. This implies
that $\Re (\alpha \bar\beta) = 0 $ and as follows $ \arg \alpha = \arg \beta + \frac{\pi}{2}    $. Solving normalization condition we find
$\alpha = \cos \gamma e^{i \arg \alpha}$,  $\beta = i\sin \gamma e^{i \arg \alpha}$. Then, the superposition state (\ref{superposition}) becomes
$$  \alpha |+\rangle + \beta |-\rangle = e^{i \arg \alpha} \left[  \frac{e^{-i\gamma} |0\rangle + e^{i\gamma} |1\rangle}{\sqrt{2}}         \right] =
     e^{i \arg \alpha- i\gamma} \left[  \frac{|0\rangle + e^{2i\gamma} |1\rangle}{\sqrt{2}}         \right].                                 $$
     This state, up to the global phase is in the form of the quantum coin (\ref{quantumcoin}), corresponding to rotation of Hadamard states along the equatorial circle.

 \section{Quantum coin flipping and measurement}
 
 The flipping of quantum coin is an application of the $X$ gate on the "heads" state $|0\rangle$ and the "tails" state $|1\rangle$. 
 If the coin is initialized in the  "heads" state $|0\rangle$, then by applying the Hadamard gate the quantum computer produces the state
 $|+ \rangle = H |0\rangle$. More generic quantum coin state $|\varphi\rangle = R(\varphi) |+\rangle = R(\varphi) H|0\rangle$ in (\ref{quantumcoin}) can be obtained by application of phase gate $R(\varphi)$ to state $|+\rangle$ as in (\ref{rotation}).

The measurement $M$ of quantum coin state,
$$ |+\rangle = \frac{1}{\sqrt{2}}( |0\rangle +  |1\rangle)  \hskip 0.5cm \line(1,0){50}\fbox{\rule[-.3cm]{0cm}{1cm} M}  \line(1,0){50}  \hskip 0.5cm |i\rangle,\,\,\,\,i=0,1 $$
gives states $|0\rangle$ or $|1\rangle$ with equal probabilities $p_0 = p_1 = \frac{1}{2}$. This is equivalent to the fair coin tossing.

\subsubsection{Universality of n-qubit measurement}

The measurement gate $M$  is the universal one qubit measurement gate.
It means that the measurement gate for an arbitrary $n$-qubit  state can be realized by applying 1-qubit
measurement gate $M$ to each of the $n$-qubits \cite{mermin}.
This becomes clear from the  generalized Born rule, applied to an arbitrary $n+1$-qubit state, which  can be represented in the form
\begin{equation}
|\psi \rangle_{n+1} = c_0 |0\rangle |\phi_0\rangle_n + c_1 |1\rangle |\phi_1\rangle_n,
\end{equation}
where states $|\phi_0\rangle_n$ and $ |\phi_1\rangle_n$ are normalized states, but not necessary orthogonal.
If, by applying the one-qubit M gate, one measures only single qubit (getting $|0\rangle$ with $p_0$ and $|1\rangle$ with $p_1$),
then the $n+1$-qubit state becomes the product state $|0\rangle |\phi_0\rangle_n$ or $|1\rangle |\phi_1\rangle_n$.
Applying this rule $n$-times, reduces the process of measurement to application of multiple copies of a single elementary hadware - the one-qubit measurement gate.

\subsection{Duplicated  states and Fibonacci numbers}
Here we count probability of measurement the quantum coin states in $n-$ trials, to get consequtive pattern of the states $|1 \rangle$ only in last two final measurements.
By generalized Born rule and universality of $n$-qubit measurement gate, we formulate problem in terms of generic $n$-qubit states.
This can be done by applying n-times the one qubit measurement gate $M$  to the Hadamard state. The results of  these measurements can be ordered as n-qubit computational state, and the first question is how many n-qubit states of following form exist:
\begin{equation} \underbrace{
|*\rangle \otimes  |*\rangle \otimes  ... |*\rangle }_{n-2}\otimes  \underbrace{|1 \rangle \otimes |1 \rangle }_2 \equiv \underbrace{|*\rangle |*\rangle   ... |*\rangle  |1 \rangle  |1 \rangle  \label{compbasis2}}_n
\end{equation}
where in the first $n-2$ measurements, the state $|1 \rangle$ can appear only once at most.
This is equivalent to ask, how many computational basis $n$-qubit states of the form (\ref{compbasis2}) exist.
Let us denote this number as $A_n$. By direct computation it is easy to get first few values
$A_2 =1$, $A_3 =1$, $A_4 =2$, $A_5 = 3$ etc.  It is noted that this problem is similar to the classical Fibonacci problem with adult and young rabbits. The state $|0 \rangle$ is associated with adult rabbit, and allows the following state from the left to be or adult state $|0\rangle$ or the young state $|1\rangle$. But the young state $|1\rangle$ can not be followed by the young state $|1\rangle$, but only by adult state $|0\rangle$
(the young becomes adult). The remarkable relation between rabbit problem and computational quantum states is shown by Fibonacci tree for $n$-qubit states in Figure 1.  

\begin{center}
 \tikzstyle{level 1}=[level distance=20mm, sibling distance=50mm]
\tikzstyle{level 2}=[level distance=20mm, sibling distance=50mm]
\tikzstyle{level 3}=[level distance=20mm, sibling distance=30mm]
\tikzstyle{level 4}=[level distance=20mm,  sibling distance=10mm]
\begin{tikzpicture}[grow=left,->,>=angle 60]
 \node {$|1_n \rangle$}
    child {node {$|1_{n-1}\rangle$}
    child {node {${ |0_{n-2}\rangle}$}
      child {node {${ |0_{n-3}\rangle}$}
       child {node {$|0\rangle$}
         child {node{$|0\rangle$}}  child {node{$|1\rangle$}}
      }
       child {node{$|1\rangle$}
         child {node{$|0\rangle$}}
      }
    }
     child {node {$|1\rangle$}
       child {node{${ |0_{n-4}\rangle}$}
         child {node{$|0\rangle$}}  child {node{$|1\rangle$}}
      }
    }
}
    };

\end{tikzpicture}

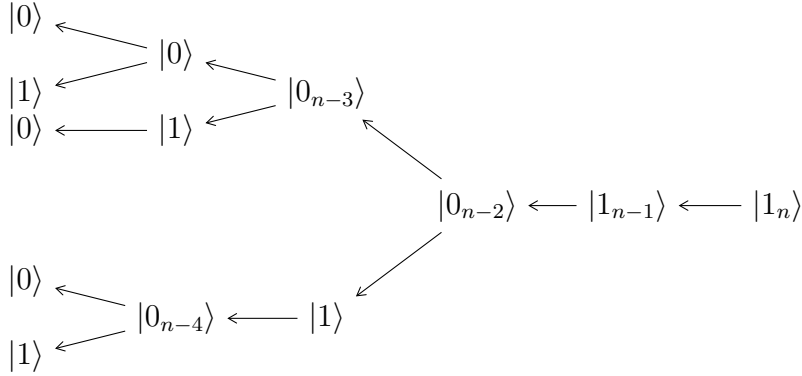
Figure 1. Fibonacci tree of states
\end{center}

 By analysing this Fibonacci tree, we notice that number $A_n$ (number of allowed states) is equivalent to the number of different paths (of length $n$) in this tree.
Starting from branching point at state $|0_{n-2}\rangle$  (Figure 1) with number $A_n$, we see that it is the sum of paths from branching state $|0_{n-3}\rangle$ with number $A_{n-1}$,
and from branching state $|0_{n-4}\rangle$ with number $A_{n-2}$.
This why 
the number of paths satisfies the following recursion formula
\begin{equation}
A_n = A_{n-1} + A_{n-2}. \label{AFibonaccirecursion}
\end{equation}
This formula is the determining relation for the Fibonacci sequence of numbers. The Fibonacci numbers $1,1,2,3,5,8,... $, are defined by the recursion formula
\begin{equation}
F_n = F_{n-1} + F_{n-2}, \label{Fibonaccirecursion}
\end{equation}
with initial values $F_0 = 0$,  $F_1 = 1$.
Comparing two sequences we have for number of states (\ref{compbasis2}) following formula
\begin{equation}
A_n = F_{n-1}, \,\,\,n = 2,3,...
\end{equation}

\subsubsection{Hadamard quantum coin measurement}

The measurements of quantum coin (\ref{quantumcoin}) give states $|0\rangle $ or $|1\rangle $ with equal probabilities $p_0 = p_1 = \frac{1}{2}$.
 Therefore, probability to have $n$-qubit
configuration (\ref{compbasis2}) is the product
\begin{equation}
P_n = A_n \,\frac{1}{2^n} = \frac{F_{n-1}}{2^n}, \,\,\,n = 2,4,....\label{PFibonacci}
\end{equation}
 Probabilities $P_n$ itself, satisfy recursion formula for the generalized Fibonacci numbers
 \begin{equation}
P_n = \frac{1}{2}P_{n-1} + \frac{1}{2^2}P_{n-2}, \label{PFibonaccirecursion}
\end{equation}
with initial values $P_1 = 0, P_2 = \frac{1}{2^2}$.    First few numbers here are $P_3 = \frac{1}{2^3}$ , $P_4 = \frac{2}{2^4}$, $P_5 = \frac{3}{2^5}$, $P_6 = \frac{5}{2^6}$, etc.
These probabilities can be calculated directly from Fibonacci tree in Figure 1, by attaching to every state the probability $\frac{1}{2}$.

\subsection{Triplicated states and Tribonacci numbers}

Now we count probability of measurement quantum coin states in $n-$ trials, to get repeated pattern of the states $|1 \rangle$ only in the last three measurements.
We can order results of  these measurements as n-qubit state, and ask how many n-qubit states of following form exist:
\begin{equation} \underbrace{
|*\rangle \otimes  |*\rangle \otimes  ... |*\rangle}_{n-3} \otimes  \underbrace{|1 \rangle \otimes |1 \rangle \otimes |1 \rangle}_3 \equiv \underbrace{|*\rangle |*\rangle   ... |*\rangle  |1 \rangle  |1 \rangle  |1 \rangle }_n\label{compbasis3}
\end{equation}
where in the first $n-3$ measurements, the state $|1 \rangle$ can appear only twice at most.
This is equivalent to ask, how many computational basis $n$-qubit states of the form (\ref{compbasis3}) exist. We call these states as allowed states.
Denoting this number as $A_n$, by direct computation we get first few values
$A_3 =1$, $A_4 =1$, $A_5 =2$, $A_6 = 4$ etc.   The set of allowed $n$-qubit states is shown in Figure 2,
which we call the Tribonacci tree for computational basis states.

\begin{center}
\tikzstyle{level 1}=[level distance=15mm, sibling distance=30mm]
\tikzstyle{level 2}=[level distance=20mm, sibling distance=30mm]
\tikzstyle{level 3}=[level distance=20mm, sibling distance=30mm]
\tikzstyle{level 4}=[level distance=15mm, sibling distance=35mm]
\tikzstyle{level 5}=[level distance=20mm, sibling distance=30mm]
\tikzstyle{level 6}=[level distance=20mm,  sibling distance=20mm]
\begin{tikzpicture}[grow=left,->,>=angle 60]
 \node {$|1_n \rangle$}
    child {node {$|1_{n-1}\rangle$}
    child {node {$|1_{n-2}\rangle$}
    child {node {${ |0_{n-3}\rangle}$}
      child {node {${ |0_{n-4}\rangle}$}          
       child {node {$|0\rangle$}                                 
         child {node{$|0\rangle$}}  child {node{$|1\rangle$}}
      }                                                                           
       child {node{$|1\rangle$}                                                           
         child {node{$|0\rangle$}}   child {node{$|1\rangle$}}
      }                                                                                                     
    }                                                  
     child {node {$|1\rangle$}    child {node{${ |0_{n-5}\rangle}$}   child {node{$|0\rangle$}}   child {node{$|1\rangle$}}
 }                                                          
       child {node{$|1\rangle$}
         child {node{${ |0_{n-6}\rangle}$}}
     }  
    }                                                                                                     
}       
}        
    };  

\end{tikzpicture}


Figure 2. Tribonacci tree of states
\end{center}

As in the above case, the number $A_n$ is equivalent to the number of different paths (of length $n$) in this tree.
Starting from branching point at state $|0_{n-3} \rangle$ (Figure 2) we see that number $A_n$ is the sum of paths at state  $|0_{n-4} \rangle$ with number $A_{n-1}$, $|0_{n-5} \rangle$ with number $A_{n-2}$
and $|0_{n-6} \rangle$ with number $A_{n-3}$. Then,
the number of paths satisfies the following recursion formula
\begin{equation}
A_n = A_{n-1} + A_{n-2} + A_{n-3}. \label{Atribonaccirecursion}
\end{equation}
This formula is determining relation for the Tribonacci sequence of numbers. The Tribonacci numbers $1,1,2,4,7,13,... $, are defined by the recursion formula
\begin{equation}
T_n = T_{n-1} + T_{n-2} + T_{n-3}, \label{tribonaccirecursion}
\end{equation}
with initial values $T_0 = 0$,  $T_1 = 0$, $T_2 = 1$.
Then, comparing two sequences we have the allowed number of  states (\ref{compbasis3}) in the form
\begin{equation}
A_n = T_{n-1}, \,\,\,n = 3,4,...
\end{equation}

 \subsubsection{Quantum coin measurement}

The measurements of Hadamard quantum coin (\ref{quantumcoin}) give states $|0\rangle $ or  $|1\rangle $, with equal probabilities $p_0 = p_1 = \frac{1}{2}$.
 Therefore, probability to have
configuration (\ref{compbasis3}) is the product
\begin{equation}
P_n = A_n \,\frac{1}{2^n} = \frac{T_{n-1}}{2^n}, \,\,\,n = 3,4,....\label{Ptribonacci}
\end{equation}
 Probabilities $P_n$ satisfy recursion formula for the generalized Tribonacci numbers
 \begin{equation}
P_n = \frac{1}{2}P_{n-1} + \frac{1}{2^2}P_{n-2} + \frac{1}{2^3}P_{n-3}, \label{Ptribonaccirecursion}
\end{equation}
with initial values $P_1 = 0, P_2 = 0, P_3 = \frac{1}{8}$.   First few numbers of this sequence are $P_4 = \frac{1}{16}$ , $P_5 = \frac{1}{16}$, $P_6 = \frac{1}{16}$, $P_7 = \frac{7}{128}$, etc.
The probabilities result also from Tribonacci tree in Figure 2, by attaching probabilities $\frac{1}{2}$ to every state.

  \subsection{N-plicated states and N-bonacci numbers}
We can generalize previous results and ask, how many $A_n$ of the $n$ qubit states ($N < n$) of the following form exist,
\begin{equation}
\underbrace{|*\rangle \otimes  |*\rangle \otimes  ... |*\rangle}_{n-N} \otimes \underbrace{ |1 \rangle \otimes |1 \rangle ...\otimes |1 \rangle}_{N} \equiv \underbrace{|*\rangle |*\rangle   ... |*\rangle |1 \rangle  |1 \rangle ... |1 \rangle}_n \label{compbasisN}
\end{equation}

For first few states we have $A_N = 1$, $A_{N+1} = 1$, $A_{N+2} = 2$, $ A_{N+3} = 4$, $A_{N+4} = 8$.
In Figure 3 we show corresponding N-bonacci tree of $n$-qubit states.

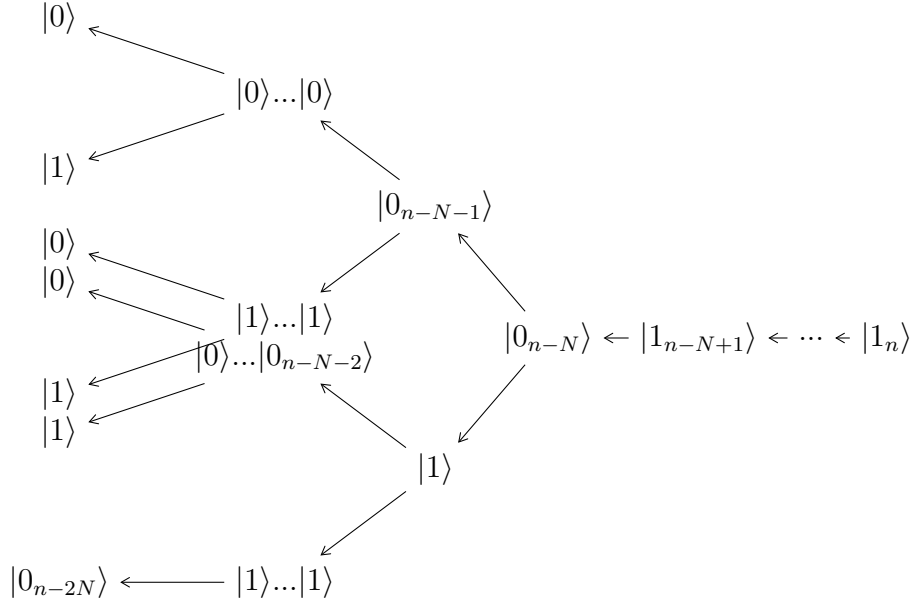
\begin{center}
 \tikzstyle{level 1}=[level distance=10mm, sibling distance=10mm]
\tikzstyle{level 2}=[level distance=15mm, sibling distance=20mm]
\tikzstyle{level 3}=[level distance=20mm, sibling distance=20mm]
\tikzstyle{level 4}=[level distance=15mm, sibling distance=35mm]
\tikzstyle{level 5}=[level distance=20mm, sibling distance=30mm]
\tikzstyle{level 6}=[level distance=30mm,  sibling distance=20mm]
\begin{tikzpicture}[grow=left,->,>=angle 60]
 \node {$|1_n \rangle$}
    child {node {$ ... $}
    child {node {$|1_{n-N+1}\rangle$}
    child {node {${ |0_{n-N}\rangle}$}
      child {node {${ |0_{n-N-1}\rangle}$}          
       child {node {$|0\rangle...|0\rangle$}                                 
         child {node{$|0\rangle$}}  child {node{$|1\rangle$}}
      }                                                                           
       child {node{$|1\rangle...|1\rangle$}                                                           
         child {node{$|0\rangle$}}   child {node{$|1\rangle$}}
      }                                                                                                     
    }                                                  
     child {node {$|1\rangle$}    child {node{$|0\rangle...{ |0_{n-N-2}\rangle}$}   child {node{$|0\rangle$}}   child {node{$|1\rangle$}}
 }                                                          
       child {node{$|1\rangle...|1\rangle$}
         child {node{${ |0_{n-2N}\rangle}$}}
     }  
    }                                                                                                     
}       
}        
    };  

\end{tikzpicture}

Figure 3. N-Bonacci tree of states
\end{center}

From this figure we can infer the recursion formula
\begin{equation}
A_{n-N} = A_{n-N-1} + A_{n-N-2} + ... + A_{n-2N},
\end{equation}
 showing  that these  numbers are expressible by the $N$-Bonacci numbers.
The $N$-Bonacci numbers $B_k$ are defined by recursion formula and initial values
\begin{eqnarray}
B_{k} = B_{k-1} + B_{k-2} + ... + B_{k-N},\\
B_0 = 0, B_1 = 0,..., B_{N-2} =0, B_{N-1} =1.
\end{eqnarray}
Then, first few $N$-Bonacci numbers are $B_N = 1$, $B_{N+1} = 2$, $B_{N+2} = 4$. Comparing these recursion formulas and initial values we conclude
that number of allowed $n$-qubit states is
\begin{equation}
  A_n = B_{n-1} ,\,\,\,\,\,\, n = N, N+1, N+2, ...
\end{equation}

\subsubsection{Hadamard quantum coin measurement}

The measurements of quantum coin (\ref{quantumcoin}) give states $|0\rangle $ or  $|1\rangle $, with equal probabilities $p_0 = p_1 = \frac{1}{2}$.
 Therefore, probability to have
configuration (\ref{compbasisN}) is the product
\begin{equation}
P_n = A_n \,\frac{1}{2^n} = \frac{B_{n-1}}{2^n}, \,\,\,n = N,N+1, ....,\label{PNbonacci}
\end{equation}
satisfying recursion formula for the generalized  $N$-Bonacci numbers
and initial values
\begin{eqnarray}
P_{n} = \frac{1}{2} P_{n-1} + \frac{1}{2^2} P_{n-2} + \frac{1}{2^3} P_{n-3} + ... + \frac{1}{2^N} P_{n-N}, \\
P_1 = P_2 = ... = P_{N-1} = 0, P_N = \frac{1}{2^N}.
\end{eqnarray}
These probabilities follow the rules of N-Bonacci tree in Figure 3, where to every state we attach the probability $\frac{1}{2}$.

\section{Generic qubit coin measurement}

As we noticed in Eq. (\ref{superposition}),  an arbitrary superposition of two quantum coins $| +\rangle$ and $| -\rangle$ is a generic qubit state. The coin, initialized in state $|0\rangle$, by applying universal one qubit gates on quantum computer, produces an arbitrary qubit state.
 The measurement of this state by the 1-qubit measurement gate M, according to the Born rule,

$$  |\psi \rangle= c_0 |0\rangle + c_1 |1\rangle  \hskip 0.5cm \line(1,0){50}\fbox{\rule[-.3cm]{0cm}{1cm} M}  \line(1,0){50}  \hskip 0.5cm |i\rangle $$
gives state $|i\rangle$, ($i=0,1$) with probability
\begin{equation}p_i = |c_i|^2 = \langle \psi |  \hat P_i | \psi \rangle, \label{genericqubit}\end{equation}
where $\hat P_i = | i \rangle \langle i |$ is projection operator $\hat P_i^2 = \hat P_i$ and $p_0 + p_1 =1$.

\subsection{Duplicated qubit measurement}

The measurement of arbitrary qubit state (\ref{qubit}) give states $|0\rangle $ or $|1\rangle $, with probabilities $p_0 = |c_0|^2$ and $p_1 = |c_1|^2$,
correspondingly. The probability of measurement of this quantum coin states in $n-$ trials, to get repeated pattern of the states $|1 \rangle$ only in the last two final measurements, we denote as $P_n$.
If we order results of  the measurements in the form of n-qubit state, then the set of allowed states is described by Fibonacci tree. 
This Fibonacci tree is shown  in Figure 1, but in this case it includes probability $p_0$ associated with every state $|0\rangle$ and $p_1$ at every state  $|1\rangle$.  The first few probabilities are
$P_2 = p_1^2$, $P_3 = P_4 = p_0 p_1^2$. The probabilities $P_n$ are given by Fibonacci polynomials in variables,  $p_0 = 1 - p_1$ and  $p_1$, satisfying recursion formula
 \begin{equation}
P_n = p_0 (P_{n-1} +  p_1 P_{n-2}), \label{anyPFibonaccirecursion}
\end{equation}
and initial values, $P_1 = 0$, $P_2 = p^2_1$.
For one-qubit state on the Bloch sphere $| \theta, \varphi\rangle = \cos \frac{\theta}{2} |0\rangle + \sin\frac{\theta}{2} e^{i \varphi} |1\rangle$, the probabilities are $p_0 = \cos^2 \frac{\theta}{2} $ and
$p_1 = \sin^2 \frac{\theta}{2}$. Then, the probabilities $P_n$ become polynomials in $\cos \theta$. The first few polynomials are 
$P_2 = \frac{1}{4} (1 - \cos \theta)^2$, $P_3 = P_4 = \frac{1}{8}(1 - \cos \theta)(1 - \cos^2 \theta )$.

\subsection{Triplicated qubit measurement}

 We denote as $P_n$ the probability of measurement of qubit coin states in $n-$ trials, with triplicated pattern of the states $|1 \rangle$ appearing only in the last three final measurements.
If we order results of  the measurements in the form of n-qubit state, then the set of allowed states is described by Tribonacci tree, shown  in Figure 2, where with every state $|0\rangle$ is associated probability $p_0$ and with state $|1\rangle$, probability $p_1$. The first few probabilities are $P_3 = p_1^3$,
$P_4 = P_5 = P_6 = p_1^3 p_0$, $P_7 = p_1^3 p_0 (1-p_1^3)$, etc. It is not difficult then to establish that these probabilities are given by Tribonacci polynomials in terms of probabilities $p_0 = 1 - p_1$, $p_1$, with recursion formula
 \begin{equation}
P_n = p_0 (P_{n-1} +  p_1 P_{n-2} + p_1^2 P_{n-3}), \label{anyPtribonaccirecursion}
\end{equation}
and initial values, $P_1 = P_2 = 0$, $P_3 = p^3_1$.
For one qubit state on Bloch sphere, the probabilies $P_n(\cos \theta)$ are polynomials in $\cos \theta$ and first few ones are
$P_3 = \frac{1}{8}(1 - \cos \theta)^3 $, $P_4 = P_5 = P_6 = \frac{1}{16} (1 - \cos \theta)^3 (1 + \cos \theta)$.

\subsection{N-plicated qubit measurement}

The recursion formula and initial values for probabilities are
\begin{eqnarray}
P_n = p_0 (P_{n-1} + p_1 P_{n-2} + ...+ p_1^{N-1} P_{n-N}), \\
P_1 = P_2 = ... = P_{N-1} = 0, P_N = p_1^N.
\end{eqnarray}
The first few polynomials $P_N = p_1^N$, $P_{N+1} = P_{N+2} = ... P_{2N} = p_0 p_1^N$,
and
\begin{equation}
P_{2N+1} = p_0^2 p_1^N (1 + p_1 + p_1^2 + ... + p_1^{N-1}) = p_0 p_1^N (1 - p^N_1),
\end{equation}
are polynomials on the Bloch sphere
$P_N = \frac{1}{2^N} (1 - \cos \theta)^N$, $P_{N+1} = ... = P_{2N} = \frac{1}{2^{N+1}}
 (1-\cos \theta)^N (1 + \cos \theta)$.
The polynomials correspond to addition of different paths on $N$-Bonacci tree (Figure 3), by attaching probability $p_0$ to every state $|0\rangle$ and $p_1$ to every state $|1\rangle$.

\section{Arbitrary position of repeated states}

In previous Section we described situation with consequtive states $|1\rangle$ in the last positions of $n$ qubit state. 
Now we are looking for probabilities to have consequtive states $|1\rangle$ in arbitrary position.
\subsection{Arbitrary position of duplicated states}
Here we treat the generic case of an arbitrary position of duplicated states in Hadamard coin states.
If duplicated $|1\rangle |1 \rangle$ states in $n$-qubit states appear only at the end positions $n-1$ and $n$,
\begin{equation} \leftarrow \underbrace{
|*\rangle \otimes  |*\rangle \otimes  ... |*\rangle}_{n-2} \otimes  \underbrace{|1 \rangle \otimes |1 \rangle }_2 \equiv \leftarrow \underbrace{|*\rangle |*\rangle   ... |*\rangle  |1 \rangle  |1 \rangle }_n,\hskip1cm A_n = F_{n-1}
\label{left}
\end{equation}
the Fibonacci tree is growing to the left and the number of allowed states is $A_n = F_{n-1}$.
If these states appear only in the first two positions $n=1$ and $n=2$, 
\begin{equation} \rightarrow \underbrace{|1 \rangle \otimes |1 \rangle }_2  \otimes  \underbrace{
|*\rangle \otimes  |*\rangle \otimes  ... |*\rangle}_{n-2}   \equiv \rightarrow \underbrace{|1 \rangle  |1 \rangle|*\rangle |*\rangle   ... |*\rangle   }_n, \hskip1cm A_n = F_{n-1}
\label{right}
\end{equation}
the Fibonacci tree grows to the right and the number of allowed states is the same $A_n = F_{n-1}$.

Now, we consider the general case of allowed states with only position for $|1\rangle |1 \rangle$ states 
at $k$ and $k+1$, where $k=1,2,...,n-1$,

 \begin{equation}\leftarrow  \underbrace{
|*\rangle \otimes  |*\rangle \otimes  ... |*\rangle}_{k-1} \otimes  \underbrace{|1 \rangle \otimes |1 \rangle }_{k \hskip0.7cm k+1} \otimes \underbrace{
|*\rangle \otimes  |*\rangle \otimes  ... |*\rangle}_{n-k-1} \rightarrow.
\label{k}
\end{equation}
The number of states is determined now by two Fibonacci trees, one is $k+1$ qubit tree growing to the left, with number of allowed states $F_k$ and another one is the tree of $n-k+1$ qubits, growing to the right, with number of the states $F_{n-k}$. Then, the total number of allowed states $A_n = F_k \cdot F_{n-k}$.
From this we conclude that for $n$-qubit state, the probability to have duplicated $|1\rangle |1\rangle$ state at position $k$, $k+1$ only, is
\begin{equation} 
P_{n,k} = \frac{F_k \cdot F_{n-k}}{2^n} = \frac{L_n - (-1)^k L_{n-2k}}{5\cdot  2^n},\label{kpositionP}
\end{equation}
where $k=1,2,...,n$, $L_n$ are Lucas numbers and we have used identity:
\begin{equation}
F_m \cdot F_n = \frac{L_{m+n} - (-1)^n L_{m-n}}{5}.
\end{equation}
The number of $n$-qubit states in which $|1\rangle |1\rangle$ states appear just once, independently of the position is
\begin{equation}
\sum^{n-1}_{k=1} F_k \cdot F_{n-k} = \frac{n \, L_n - F_n}{5}.\label{arbitraryposition}
\end{equation}
The corresponding probability to have this states once, but anywhere is
\begin{equation} 
\sum^{n-1}_{k=1} P_{n,k} = \sum^{n-1}_{k=1}\frac{F_k \cdot F_{n-k}}{2^n} = \frac{n \, L_n - F_n}{5\cdot 2^n}.\label{arbitrarypositionprobability}
\end{equation}

\subsection{Triplicated states at arbitrary position}

For $n$-trials with triplicated state $|1\rangle |1 \rangle |1 \rangle$ at positions $n-2$, $n-1$, $n$,
the Tribonacci tree is growing in the left direction and number of allowed states is $A_n = T_{n-1}$, $n=3,4,...$.
The same number of states is for the triplicated state at position $1$, $2$, $3$, with Tribonacci  states tree, growing to the right. Now, if the triplicated state is  located at arbitrary positions $k$, $k+1$, $k+2$, then the number of allowed states is product of number $T_{k+1}$ of states for left Tribonacci tree, and  
$T_{n-k}$ for the right Tribonacci tree. This gives number of allowed states at arbitrary position $k$ as
\begin{equation}
A_{n, k} = T_{k+1} T_{n-k},
\end{equation}
where $k =1, 2,...,n-2$. The probability of this configurartion is 
\begin{equation}
P_{n,k} = \frac{T_{k+1} T_{n-k}}{2^n}
\end{equation}
If position of triplicated states is not fixed, then the number of states and corresponding probabilities are
\begin{equation}
A_{n, k} = \sum^{n-2}_{k=1} T_{k+1} T_{n-k}, \,\,\,\,P_{n, k} = \sum^{n-2}_{k=1}
 \frac{T_{k+1} T_{n-k}}{2^n}
\end{equation}

\subsection{The N-plicated states in arbitrary position}

For the $N$-plicated states $|1 \rangle...|1 \rangle$ at arbitrary position $k$ we have number of allowed states and corresponding probabilities in terms of $N$-Bonacci numbers
\begin{equation}
A_{n, k} = B_{k+N-2} B_{n-k},\,\,\,\,\,P_{n,k} = \frac{B_{k+N-2} B_{n-k}}{2^n},
\end{equation}
where $k =1,2,...,n-N+1$.

\section{Generating function and Golden Ratio}

\subsection{Generating function for  duplicated probability}
For duplicated states of Hadamard quantum coin, the generating function of probabilities  (\ref{PFibonacci})    is
\begin{equation}
g(x) = \sum^\infty_{n=2} P_n \,x^{n-2} = \sum^\infty_{n=0} P_{n+2}\, x^{n} = \sum^\infty_{n=0}\frac{F_{n+1}}{2^{n+2}} x^n=\frac{1}{4 - 2 x -  x^2}.
\end{equation}
This can be shown by using recursion formula (\ref{Fibonaccirecursion}) or another way, by using Binet representation for Fibonacci numbers and geometric series.
For $x=1$ it gives identity
\begin{equation}
\sum^\infty_{n=0} P_n =   \sum^\infty_{n=0}\frac{F_{n+1}}{2^{n+2}}   = 1, \label{unit}
\end{equation}
The identity is the completeness relation for duplicated states (\ref{compbasis2}) and it shows that duplicated state will appear once in the set of all numbers $n$ trials. 
This relation allows to introduce information characteristic of the states by the Shannon entropy (\ref{shannon})
and orreponding qubit states, realized by random walk in Fock space of computational states. 

\subsubsection{Golden ratio in computational states}

For duplicated $n$-qubit states (\ref{compbasis2}), the number of states is $A_n = F_{n-1}$, and for duplicated $n+1$-qubit states it is $A_{n+1} = F_n$. The ratio of these numbers in the limit $n \rightarrow \infty$ is the Golden ratio
\begin{equation}
\lim_{n \rightarrow \infty} \frac{A_{n+1}}{A_n} = \frac{F_n}{F_{n-1}} = \varphi.
\end{equation}

\subsubsection{Golden ratio in Hadamard coin measurements }

The probability to get duplicated states (\ref{compbasis2}) in $n$-measurements of the Hadamard coin states is 
$P_n = \frac{F_{n-1}}{2^n}$ and for $n+1$-measuremnts it is $P_{n+1} = \frac{F_{n}}{2^{n+1}}$.
The ratio of these probabilities in the limit $n \rightarrow \infty$ is half of the Golden ratio:
\begin{equation}
\lim_{n \rightarrow \infty} \frac{P_{n+1}}{P_n} = \lim_{n \rightarrow \infty} \frac{F_n}{2 F_{n-1}} = \frac{1}{2} \varphi.
\end{equation}

\subsubsection{Golden Ratio for arbitrary position}
We take $n$-qubit state with allowed number of states  $A_n$ and $n+1$-qubit state with number $A_{n+1}$ from Eq.(\ref{arbitraryposition}). The ratio of these numbers in the limit $n \rightarrow \infty$ is the Golden ratio
\begin{equation}
\frac{A_{n+1}}{A_n} = \frac{(n+1) \, L_{n+1} - F_{n+1}}{n \, L_n - F_n} \rightarrow  \varphi.
\end{equation}
The ratio of corresponding probabilities (\ref{arbitrarypositionprobability})
 to find $|1\rangle |1\rangle$ pair at arbitrary position, but just only once, in the set of $n$ and $n+1$ trials, in the limit $n \rightarrow \infty$ is half of the Golden Ratio
\begin{equation}
\frac{P_{n+1}}{P_n} = \frac{1}{2}\frac{(n+1) \, L_{n+1} - F_{n+1}}{n \, L_n - F_n} \rightarrow  \frac{1}{2} \varphi.
\end{equation}
The same limit is valid for ratio of probabilies $P_{n,k}$ in (\ref{kpositionP}) at position $k$, $k+1$, where $k=1,2,...,n-1$.

\subsubsection{Shannon entropy}

The Hadamard quantum tree in Figure 1, at every level $n$ determines probability $P_n = \frac{F_{n-1}}{2^n}$ and as we have seen, the sum of probabilities (\ref{unit}) is one, $\sum^\infty_{n=2}P_n  =1 $. The level of randomness in this distribution can be characterized by  the Shannon entropy
\begin{equation} 
S = - \sum^\infty_{n=2} P_n \log_2 P_n = - \sum^\infty_{n=2} \frac{F_{n-1}}{2^n} \log_2 \frac{F_{n-1}}{2^n}.\label{shannon}
\end{equation}
The ratio test for this series gives half of the Golden ratio
 $\lim_{n \rightarrow \infty} \frac{s_{n+1}}{s_n} = \frac{1}{2} \varphi < 1$ and convergency of this sum.
For the $n$-th term, assymptotically we have
\begin{equation}
s_n \approx n \left(  \frac{\varphi}{2} \right)^n \frac{\log_2 \frac{2}{\varphi}}{\sqrt{5}}.
\end{equation}

\subsection{Generating function for triplicated probability}

The generating function of Tribonacci numbers is defined as
\begin{equation}
g_T(x) = \sum^\infty_{n=0} T_n x^n = x^2 + x^3 + 2 x^4 + 4 x^5 + ...\label{gentribonacci}
\end{equation}
By substituting the recursion formula for Tribonacci numbers (\ref{tribonaccirecursion}) and using values $T_{-1} = 1$, $T_{-2} = -1$, $T_{-3} = 0$
we get
\begin{equation}
g_T(x) = \frac{x^2}{1 - x - x^2 - x^3}.\label{TribGen}
\end{equation}

For Hadamard quantum coin (\ref{quantumcoin}) we can define generating function of Tribonacci probabilities $P_n$ (\ref{Ptribonacci}),
\begin{equation}
g(x) = \sum^\infty_{n=0} P_{n+3}\, x^n =  \sum^\infty_{n=0} \frac{T_{n+2}}{2^{n+3}}\,x^n .\label{genPtribonacci}
\end{equation}
By using recursion formula for Tribonacci numbers we find
\begin{equation}
g(x) = \frac{1}{2^3 - 2^2 x - 2 x^2 - x^3}.\label{PTribGen}
\end{equation}
If in this formula we take $x=1$, then we find completeness  relation for probabilities
\begin{equation}
\sum^\infty_{n=3} P_n = 1.
\end{equation}
The last relation allows to introduce the Shannon entropy for triplicated states in Fock space in a similar to (\ref{shannon}) way.

\section{Projection operators and $n$-qubit states}

\subsection{Maximally random n-qubit state}

The $n$-qubit state
\begin{equation}
|\psi \rangle = \frac{1}{\sqrt{2^n}} \sum^1_{i_{n-1} ...i_1 i_0}  |i_{n-1} ...i_1 i_0\rangle = \frac{1}{\sqrt{2^n}} \sum^{2^n -1}_{i = 0}  |i\rangle \label{nmax}
\end{equation}
is maximally random state, according to the computational basis. In this case, all probabilities are equal and the Shannon entropy is maximal $S = n $. 
In Fibonacci states tree, shown in Figure 1, with every path we can associate the projection operator on corresponding computational state. Then, by adding these projection operators we have projection operator to the subspace of $n$-qubit states, determined by Fibonacci tree.
For first five qubit states we have projection operators 
\begin{eqnarray}
\hat P_2 &=& |11\rangle \langle 11|, \\
\hat P_3 &=& |011\rangle \langle 011|, \\
\hat P_4 &=& |0011\rangle \langle 0011| +|1011\rangle \langle 1011| , \\
\hat P_5 &=& |00011\rangle \langle 00011| +|01011\rangle \langle 01011| + |10011\rangle \langle 10011|.
\end{eqnarray}
For arbitrary $n$-qubit state, projection operator to Fibonacci tree is
\begin{equation}
\hat P_n = \sum^{\neq 11}_{i_{n-1},...,i_3 =0,1} | i_{n-1}...i_3 011\rangle \langle i_{n-1}...i_3 011|,\label{projection}
\end{equation}
where in summation, the duplicated states $|1\rangle |1\rangle$ are not included. Dimension of the Hilbert space, associated with $n$-qubit Fibonacci tree in Figure 1 is given by Fibonacci number $dim_F H =F_{n-1}$. For the first five qubits in the above examples we have number of dimensions correspondingly,  $ 1,1,2,3$.
Acting by projection operator(\ref{projection})  on state (\ref{nmax}) we get the state
\begin{equation}
|\phi \rangle=\hat P_n |\psi \rangle,
\end{equation}
connected with Fibonacci tree in Figure 1. The inner product of this state 
\begin{equation}
\langle \phi |\phi \rangle= \frac{F_{n-1}}{2^n}
\end{equation}
gives for corresponding normalized state
\begin{equation}
|\Phi \rangle = \frac{|\phi \rangle}{\sqrt{\langle \phi | \phi\rangle }} = \frac{1}{\sqrt{F_{n-1}}}
 \sum^{\neq 11}_{i_{n-1},...,i_3 =0,1} | i_{n-1}...i_3 011\rangle .
\end{equation}
Then, average of the projection operator describes probability of collapsing of the state (\ref{nmax}) to Fibonacci tree state
\begin{equation}
P_n = \langle \psi | \hat P_n | \psi \rangle = \frac{F_{n-1}}{2^n}.
\end{equation}
This formula coinsides with probability (\ref{PFibonacci}) of Hadamard coin measurement in $n$ trials.

Similar computations can be done for the triplicated and $N$-plicated states, associated with corresponding trees. For triplicated states we have projection operator
\begin{equation}
\hat P_n = \sum^{\neq 111}_{i_{n-1},...,i_4 =0,1} | i_{n-1}...i_4 0111\rangle \langle i_{n-1}...i_3 0111|,
\end{equation}
associated with Tribonacci tree in Figure 2. Dimension of the corresponding Hilbert space is determined by Tribonacci numbers $dim_T H = T_{n-1}$. Action of this operator to the state (\ref{nmax}), $\hat P_n |\psi\rangle$, gives the normalized state
\begin{equation}
|\Phi \rangle  = \frac{1}{\sqrt{T_{n-1}}}
 \sum^{\neq 111}_{i_{n-1},...,i_4 =0,1} | i_{n-1}...i_4 0111\rangle .
\end{equation}
Then, probability of state (\ref{PFibonacci}) collapse to the Tribonacci tree subspace is
\begin{equation}
P_n = \langle \psi | \hat P_n | \psi \rangle = \frac{T_{n-1}}{2^n}.
\end{equation}
In the general case of $N$ duplicated states, projection operator associated with $N$ Bonacci tree in  Figure 3, is 
\begin{equation}
\hat P_n = \sum^{\neq 11...1}_{i_{n-1},...,i_{N+1} =0,1} | i_{n-1}...i_{N+1} 011...1\rangle \langle i_{n-1}...i_{N+1} 011...1|.
\end{equation}
The dimension of corresponding Hilbert subspace $dim_B H = B_{n-1} $ and probability 
\begin{equation}
P_n = \langle \psi | \hat P_n | \psi \rangle = \frac{B_{n-1}}{2^n},
\end{equation}
are determined by the $N$ Bonacci numbers $B_n$.

\subsection{Arbitrary $n$-qubit state}
The above projection operators can be applied also to an arbitrary $n$-qubit state
\begin{equation}
|\Psi \rangle = \sum_{i_{n-1},...,i_{0} = 0}^1 c_{i_{n-1}...i_{0}} | i_{n-1}...i_{0} \rangle = 
\sum_{i=0}^{2^n -1} c_i |i \rangle,\label{genericnqubit}
\end{equation}
where $ \sum^{2^n -1}_{i=0} |c_i|^2 =1 $. Depending on projector, the state would be projected to subspaces, determined by Fibonacci tree (Figure 1), Tribonacci tree (Figure 2) and in general, the $N$-Bonacci tree (Figure 3). Then, the average of projection operator   $\hat P_n$   gives probability of collapse to arbitrary $N$-Bonacci state as
\begin{equation}
P_n = \langle \Psi | \hat P_n | \Psi \rangle = \sum^{\neq 11...11}_{i_{n-1},...,i_{N+1} = 0,1}
 |c_{i_{n-1}...i_{N+1} 0 11...11} |^2.
\end{equation}

\subsection{Decimal form of computational states}

The second sum in (\ref{genericnqubit}) represents expansion of $n$-qubit state in computational basis, counted in decimal base. By using this form, we can associate with Fibonacci tree a sequence of numbers.
To specify the same numbers, but with different number of qubits, we will use following notation. Binary numbers $11$, $011$ and $0011$ determine the same decimal number 3, this is why for corresponding computational  states we denote number of qubits by subsript, $|11\rangle = |3\rangle_2$, $|011\rangle = |3\rangle_3$, $|0011\rangle = |3\rangle_4$. 
In Figure 4 we show Fibonacci tree for these states. 

\begin{center}
 \tikzstyle{level 1}=[level distance=20mm, sibling distance=50mm]
\tikzstyle{level 2}=[level distance=20mm, sibling distance=50mm]
\tikzstyle{level 3}=[level distance=20mm, sibling distance=30mm]
\tikzstyle{level 4}=[level distance=20mm,  sibling distance=10mm]
\begin{tikzpicture}[grow=left,->,>=angle 60]
 \node {$|1 \rangle_1$}
    child {node {$|3\rangle_2$}
    child {node {${ |3\rangle_3}$}
      child {node {${ |3\rangle_4}$}
       child {node {$|3\rangle_5$}
         child {node{$|3\rangle_6$}}  child {node{$|35\rangle_6$}}
      }
       child {node{$|19\rangle_5$}
         child {node{$|19\rangle_6$}}
      }
    }
     child {node {$|11\rangle_4$}
       child {node{${ |11\rangle_5}$}
         child {node{$|11\rangle_6$}}  child {node{$|43\rangle_6$}}
      }
    }
}
    };

\end{tikzpicture}

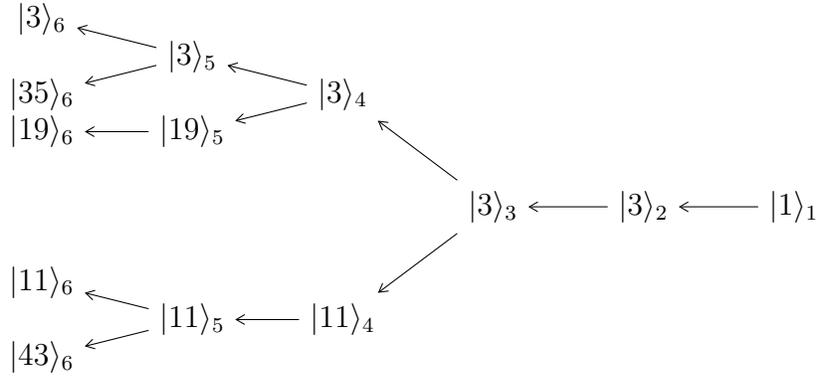
Figure 4. Fibonacci tree of decimal states
\end{center}

This tree determines the sequence of numbers and corresponding states. The sequence of numbers is: 3, 11, 19, 35, 43, 67, 75, 83, 131, 139, 147, 163, 171, etc. If we count the number of these numbers $A_n < 2^n$ for given $n$, then it is equal to Fibonacci number $A_n = F_{n-1}$. 
The set of corresponding $n$-qubit states is shown in Figure 4:       $\{|3\rangle_2 \}$ , $\{ |3\rangle_3\}$, $\{ |3\rangle_4,  |11\rangle_4 \}$, $\{|3\rangle_5,  |11\rangle_5, |19\rangle_5\}$,
$\{   |3\rangle_6,  |11\rangle_6   , |19\rangle_6,   |35\rangle_6,  |43\rangle_6  \} $ with number of states, $1, 1, 2, 3, 5$.
 The number of these $n$-qubit states from Fibonacci tree is Fibonacci number $F_{n-1}$. 
If instead of last two positions we count duplicated states from the first position or from arbitrary position $k$, then for every $k$ we will have specific sequence of numbers, depending on $k$.

This counting can be extended for triplicated and generic  $N$-plicated states as well. The Tribonacci tree in Figure 5 determines the set of numbers: 7, 23, 39, 55, 71, 87, 103, 135, 151, 167, 183, 199, 215, etc.
The number of these numbers $A_n < 2^n$ is equal to Tribonacci numbers $A_n = T_{n-1}$.
The Tribonacci tree in Figure 5 show the set of corresponding computational states:$\{|7\rangle_3 \}$ , $\{ |7\rangle_4\}$, $\{ |7\rangle_5,  |23\rangle_5 \}$, $\{|7\rangle_6,  |23\rangle_6, |39\rangle_6,
|55\rangle_6     \}$,
$\{   |7\rangle_7,  |23\rangle_7  , |39\rangle_7, $ $  |55\rangle_7,  |71\rangle_7 , |87\rangle_7, |103\rangle_7 \} $ with number of states, $1, 1, 2, 4, 7$.
The number of these $n$-qubit states is equal to Tribonacci numbers $T_{n-1}$.

\begin{center}
\tikzstyle{level 1}=[level distance=15mm, sibling distance=30mm]
\tikzstyle{level 2}=[level distance=20mm, sibling distance=30mm]
\tikzstyle{level 3}=[level distance=20mm, sibling distance=30mm]
\tikzstyle{level 4}=[level distance=15mm, sibling distance=35mm]
\tikzstyle{level 5}=[level distance=20mm, sibling distance=30mm]
\tikzstyle{level 6}=[level distance=20mm,  sibling distance=20mm]
\begin{tikzpicture}[grow=left,->,>=angle 60]
 \node {$|1 \rangle_1$}
    child {node {$|3\rangle_2$}
    child {node {$|7\rangle_3$}
    child {node {${ |7\rangle_4}$}
      child {node {${ |7\rangle_5}$}          
       child {node {$|7\rangle_6$}                                 
         child {node{$|7\rangle_7$}}  child {node{$|71\rangle_7$}}
      }                                                                           
       child {node{$|39\rangle_6$}                                                           
         child {node{$|39\rangle_7$}}   child {node{$|103\rangle_7$}}
      }                                                                                                     
    }                                                  
     child {node {$|23\rangle_5$}    child {node{${ |23\rangle_6}$}   child {node{$|23\rangle_7$}}   child {node{$|87\rangle_7$}}
 }                                                          
       child {node{$|55\rangle_6$}
         child {node{${ |55\rangle_7}$}}
     }  
    }                                                                                                     
}       
}        
    };  

\end{tikzpicture}


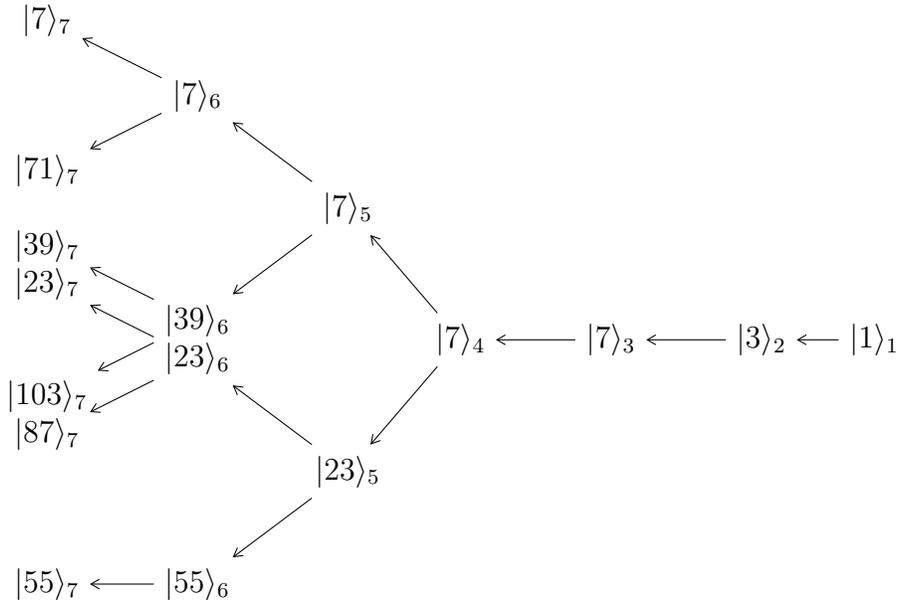
Figure 5. Tribonacci tree of decimal states
\end{center}

\section{Quantum qutrit coin}

The above theory can be extended to more general quantum coins, related with qutrit and generic qudit units of quantum information. The corresponding states are determined by generalized Fibonacci, Tribonacci and 
$N$-Bonacci numbers.

\subsection{Duplicated States}
The state of the coin is superposition of three basis states
\begin{equation}
|\psi \rangle = \frac{1}{\sqrt{3}} (|0 \rangle + |1 \rangle + |2\rangle).
\end{equation}
Every measurement of this coin produces basis states with equal probabilities $\frac{1}{3}$, so that this state is the most random qutrit state in computational basis with base 3.
In the set of $n$ trials we count aloowed states with duplicated last two states
\begin{equation} \underbrace{
|*\rangle \otimes  |*\rangle \otimes  ... |*\rangle }_{n-2}\otimes  \underbrace{|1 \rangle \otimes |1 \rangle }_2 \equiv \underbrace{|*\rangle |*\rangle   ... |*\rangle  |1 \rangle  |1 \rangle ,  \label{qutritcompbasis2}}_n
\end{equation}
where, in contrast to qubit case, we have three states $|*\rangle = |0\rangle, |1\rangle, |2\rangle$.
To count the number of allowed states, we can analyze the corresponding states tree, in a similar way as in Figure 1. But in this case the tree is duplicating its branches, so that number of states $B_n = E_{n-2}$ , where
numbers $E_n$ are the generalized Fibonacci numbers with recursion formula
\begin{equation}
E_n = 2 (E_{n-1} + E_{n-2}),
\end{equation}
and initial conditions $E_0 =0, E_1 =1$. The first few numbers are 0, 1, 2, 6, 16, 44.
Probability of duplicated states $|1\rangle |1\rangle$ in $n-$trials for the qutrit quantum coin is
\begin{equation}
P_n = \frac{E_{n-1}}{3^n}.
\end{equation}
The recursion formula for these probabilities is
\begin{equation}
P_{n+1} = \frac{2}{3} P_n + \frac{2}{3^2} P_{n-1},
\end{equation}
with initial values $P_2 = \frac{1}{3^2}, P_3 = \frac{2}{3^3}$. These probabilities are countable from corresponding doubled Fibonacci tree with probabilities $\frac{1}{3}$, associated with every state.

\subsection{Triplicated States}

\subsubsection{Maximally Random Coin}
The number $B_n = E_{n-1}$ of triplicated states 
\begin{equation} \underbrace{
|*\rangle \otimes  |*\rangle \otimes  ... |*\rangle}_{n-3} \otimes  \underbrace{|1 \rangle \otimes |1 \rangle \otimes |1 \rangle}_3 \equiv \underbrace{|*\rangle |*\rangle   ... |*\rangle  |1 \rangle  |1 \rangle  |1 \rangle }_n\label{qutritcompbasis3}
\end{equation}
 where $|*\rangle = |0\rangle, |1\rangle, |2\rangle$, is written in terms of the generalized Tribonacci numbers, with  recursion formula
\begin{equation}
E_n = 2 (E_{n-1} + E_{n-2} + E_{n-3}),
\end{equation}
and initial values $E_0 = 0, E_1 = 0, E_2 = 1$. The first five numbers are $0, 0, 1, 2, 6, 18$. Probabilities to have triplicated state at the end of $n$ trials 
\begin{equation}
P_n = \frac{E_{n-1}}{3^n}
\end{equation}
is subject to recursion formula
\begin{equation}
P_n = \frac{2}{3} (P_{n-1} + \frac{1}{3}P_{n-2} + \frac{1}{3^2}P_{n-3}),
\end{equation}
and initial values $P_1 = 0, P_2 = 0, P_3 = \frac{1}{3^3}$. It is also countable from correspondiing doubled Tribonacci tree by assigning probability $\frac{1}{3}$ to every state.

\subsubsection{Arbitrary Qutrit Coin}

For arbitrary qutrit coin
\begin{equation}
|\psi \rangle = c_0 |0\rangle + c_1 |1\rangle + c_2 |2\rangle
\end{equation}
probabilities are $p_i = |c_i|^2$, $p_0 + p_1 + p_2 =1$. For triplicated state in $n$ trials we have probabilities $P_n$, as generalized Tribonacci polinomials in $p_1$,
\begin{equation}
P_{n+1} = (1-p_1) (P_n + p_1 P_{n-1} + p_1^2 P_{n-2}),
\end{equation}
were $P_1 = P_2 = 0$, $P_3 = p^3_1$. To count these probabilities from doubled Tribonacci tree, we have to associate probabilities: $p_0$ with state $|0 \rangle$, $p_1$ with state $|1 \rangle$
and $p_2$ with state $|2 \rangle$.

\section{Quantum Qudit Coin}

\subsection{Duplicated States}

The state of the qudit coin is superposition of $d$ basis states
\begin{equation}
|\psi \rangle = \frac{1}{\sqrt{d}} (|0 \rangle + |1 \rangle + ... + |d-1\rangle).
\end{equation}
Measurement on this coin gives the basis states with equal probabilities $\frac{1}{d}$ and the state is maximally random state in computational qudit  basis with base $d$.  The measurement in $n$-trials gives the state
\begin{equation} \underbrace{
|*\rangle \otimes  |*\rangle \otimes  ... |*\rangle }_{n-2}\otimes  \underbrace{|1 \rangle \otimes |1 \rangle }_2 \equiv \underbrace{|*\rangle |*\rangle   ... |*\rangle  |1 \rangle  |1 \rangle ,  \label{quditcompbasis2}}_n
\end{equation}
where $|*\rangle = |0\rangle, |1\rangle,...,|d-1\rangle$. The number of allowed states 
$B_n = D_{n-1}$
is expressed by the generalized Fibonacci numbers $D_n$, with recursion relation
\begin{equation}
D_n = (d-1) (D_{n-1} + D_{n-2}),
\end{equation}
and initial values $D_0 =0, D_1 =1$. First few numbers are polynomials in $d$: $D_0 = 0, D_1 =1, D_2 = d-1, D_3 = (d-1)d, D_4 = (d-1)^2 (d+1), D_5 = (d-1)^2 (d^2 + d -1)$. For probability of qudit coin in $n$ trials we have
\begin{equation}
P_n = \frac{D_{n-1}}{d^n}.
\end{equation}
The recursion formula for these probabilities is
\begin{equation}
P_{n+1} = \frac{d-1}{d} P_n + \frac{d-1}{d^2} P_{n-1},
\end{equation}
with initial values $P_2 = \frac{1}{d^2}, P_3 = \frac{d-1}{d^3}$.

For generic qudit coin state
\begin{equation}
|\psi \rangle = c_0 |0 \rangle + c_1 |1 \rangle + ... + c_{d-1}|d-1\rangle \label{qdit}
\end{equation}
probabilities to measure basis states are $p_i = |c_i|^2$, $i=0,1,...,d-1$ and $\sum^{d-1}_{i=0} p_i =1$. 
Probability $P_n$ to measure duplicated state $|1\rangle |1\rangle$ only at the end of $n$ trials, satisfies the recursion formula
\begin{equation}
P_n = (1-p_1) P_{n-1} + (1 - p_1) p_1 P_{n-2},
\end{equation}
with initial values $P_2 = p^2_1$, $P_3 = p_1^2 (1-p_1)$. In thess formulas $1- p_ 1= p_0 + p_2 +...+ p_{d-1}$. The formula can be generalized also to the case, when duplicated state at the end of $n$ trials is $|k\rangle |k \rangle$, where $k=0,1,2,..,d-1$. The probability to have thise state
\begin{equation} \underbrace{
|*\rangle \otimes  |*\rangle \otimes  ... |*\rangle }_{n-2}\otimes  \underbrace{|k \rangle \otimes |k \rangle }_2 \equiv \underbrace{|*\rangle |*\rangle   ... |*\rangle  |k \rangle  |k \rangle ,  \label{kquditcompbasis2}}_n
\end{equation}
satisfies recursion formula
\begin{equation}
P_n = (1-p_k) P_{n-1} + (1 - p_k) p_k P_{n-2},
\end{equation}
with initial values $P_2 = p^2_k$, $P_3 = p_k^2 (1-p_k)$ and $1-p_k = p_0 +...+p_{k-1} + p_{k+1}+...+p_{d-1}$.

\subsection{Triplicated States}

For maximally random coin, number of allowed states $B_n = E_{n-1}$, where $E_n$ are generalized Tribonacci numbers
\begin{equation}
E_n = (d-1) (E_{n-1} + E_{n-2} + E_{n-3}),
\end{equation}
and initial values $E_0 = 0, E_1 = 0, E_2 = 1$. First few numbers are $E_3 = d-1$, $E_4 = (d-1) d$,
$E_5 = (d-1) d^2$. For probabilities of these states it gives
\begin{equation}
P_n = \frac{E_{n-1}}{d^n}
\end{equation}
and recursion formula
\begin{equation}
P_{n+1} = \frac{d-1}{d} \left(P_{n} + \frac{1}{d} P_{n-1} + \frac{1}{d^2} P_{n-2}\right),
\end{equation}
with $P_1 = P_2 =0$, $P_3 = \frac{1}{d^3}$, $P_4 = \frac{d-1}{d^4}$.

For generic qudit coin (\ref{qdit}), probability of triplicated state $P_n$ is polynomial in $p_1$, 
\begin{equation}
P_{n+1} = (1-p_1) (P_{n} + p_1 P_{n-1} + p^2_1 P_{n-2}),
\end{equation}
with $P_1 = P_2 =0$, $P_3 = p^3_1$, $P_4 = p^3_1 (1-p_1)$, 
where $1-p_1 = p_0 +p_2 + p_3+...+ p_{d-1}$.

\subsection{N-plicated States}

For arbitrary qudit state (\ref{qdit}), probability $P_n$ to find N-plicated state
\begin{equation}
\underbrace{|*\rangle \otimes  |*\rangle \otimes  ... |*\rangle}_{n-N} \otimes \underbrace{ |1 \rangle \otimes |1 \rangle ...\otimes |1 \rangle}_{N} \equiv \underbrace{|*\rangle |*\rangle   ... |*\rangle |1 \rangle  |1 \rangle ... |1 \rangle}_n \label{quditcompbasisN}
\end{equation}
 where $|*\rangle = |0\rangle, |1\rangle,..., |d-1\rangle$, is expressed by generalized $N$-Bonacci polynomials
\begin{equation}
P_{n} = (1-p_1) (P_{n-1} + p_1 P_{n-2} + p^2_1 P_{n-3} + ... + p_1^{N-1} P_{n-N}),
\end{equation}
with $P_1 = P_2 = ... = P_{N-1} = 0$, $P_N = p^N_1$,
where $1-p_1 = p_0 +p_2 + p_3+...+ p_{d-1}$.

\section{Acknowledgements} This work was partially supported by TUBITAK grant 116F206.

\section{Conflicts of Interest}

The authors declare that they have no conflicts of interest.



\begin{thebibliography}{99}

\bibitem{p1} O. K. Pashaev and S. Nalci, Golden quantum oscillator and Binet-Fibonacci calculus, 2012\textit{J. Phys A: Math Theor} \textbf{45} 015303

\bibitem{p2} O. K. Pashaev, Quantum calculus of Fibonacci divisors and infinite hierarchy of Bosonic-Fermionic Golden quantum oscillators, 2020, arXiv:2010.12386v1, math-ph; \textit{IJGMMP} (in press)

\bibitem{demoivre} A. De Moivre,  The doctrine of chanses, 3rd edition 1756 Chelsea Publishing Co., New York (reprint) 1967
\bibitem{kennedi} S. F. Kennedy  and M. W. Stafford,  Coin flipping, dynamical systems and the Fibonacci numbers 1994 \textit{Mathematics Magazine} \textbf{67}
    380-382
\bibitem{aharonov} Y. Aharonov, L. Davidovich and N. Zagury, Quantum random walks 1993 \textit{Phys Rev A} \textbf{48} 1687-1690
\bibitem{franco} C. Di Franco , M. Mc Gettrick and T. Machida,  Alternate two-dimensional quantum walk with a single-qubit coin 2011 \textit{Phys Rev A}
    \textbf{84} 042337
\bibitem{quantumwalk} Salvador Elias Venegas-Andraca,  Quantum walks: a comprehensive review 2012 \textit{Quantum Information Processing} \textbf{}
    380-382
\bibitem{manko} V. N. Chernega, O. V. Man'ko and V. I. Man'ko, Probability representation of quantum states as a renaissance of hidden variables - God plays coins
    2019 \textit{Journal of Russian Laser Research} \textbf{40} 107-120
\bibitem{albrecht} A. Albrecht  and D. Phillips, Origin of probabilities and their application to the multiverse 2014 \textit{Phys Rev D} \textbf{90} 123514
\bibitem{deutsch} D. Deutsch, Uncertainty in quantum measurements 1983 \textit{Phys Rev Lett}
    \textbf{50} 631
\bibitem{mermin} D. N. Mermin, Quantum Computer Science, Cambridge University Press, UK, 2007

 
\end{thebibliography}
\end{document}